\documentclass[conference]{IEEEtran}
\IEEEoverridecommandlockouts
\usepackage{cite}
\usepackage{color}
\usepackage{ulem}
\usepackage{amsmath,amssymb,amsfonts}
\usepackage{algorithmic}
\usepackage{graphicx}
\usepackage{textcomp}
\usepackage{xcolor}
\def\BibTeX{{\rm B\kern-.05em{\sc i\kern-.025em b}\kern-.08em
    T\kern-.1667em\lower.7ex\hbox{E}\kern-.125emX}}
\usepackage[linewidth=0.5pt]{mdframed}
\usepackage{hyperref}[colorlinks,linkcolor=blue]
\usepackage[linesnumbered,boxed,ruled,commentsnumbered]{algorithm2e}
\begin{document}

\title{An Opponent-Aware Reinforcement Learning Method for Team-to-Team Multi-Vehicle Pursuit via Maximizing Mutual Information Indicator

 \thanks{This work was supported by the National Natural Science Foundation of China (Grant No. 62071179) and project A02B01C01-201916D2}
}

\author{\IEEEauthorblockN{Qinwen Wang,
Xinhang Li,
Zheng Yuan,
Yiying Yang, 
Chen Xu, and
Lin Zhang
}
\IEEEauthorblockA{School of Artificial Intelligence, Beijing University of Posts and Telecommunications, Beijing, China}
\IEEEauthorblockA{\{wangqinwen, lixinhang, yuanzheng, yyying, chen.xu, zhanglin\}@bupt.edu.cn}
}

\maketitle

\begin{abstract}
The pursuit-evasion game in Smart City brings a profound impact on the Multi-vehicle Pursuit (MVP) problem, when police cars cooperatively pursue suspected vehicles. Existing studies on the MVP problems tend to set evading vehicles to move randomly or in a fixed prescribed route. The opponent modeling method has proven considerable promise in tackling the non-stationary caused by the adversary agent. However, most of them focus on two-player competitive games and easy scenarios without the interference of environments. This paper considers a Team-to-Team Multi-vehicle Pursuit (T2TMVP) problem in the complicated urban traffic scene where the evading vehicles adopt the pre-trained dynamic strategies to execute decisions intelligently. To solve this problem, we propose an opponent-aware reinforcement learning via maximizing mutual information indicator (OARL$ \text{M}^2 \text{I}^2 $) method to improve pursuit efficiency in the complicated environment. First, a sequential encoding-based opponents joint strategy modeling (SEOJSM) mechanism is proposed to generate evading vehicles’ joint strategy model, which assists the multi-agent decision-making process based on deep Q-network (DQN). Then, we design a mutual information-united loss, simultaneously considering the reward fed back from the environment and the effectiveness of opponents joint strategy model, to update pursuing vehicles’ decision-making process. Extensive experiments based on SUMO demonstrate our method outperforms other baselines by 21.48$\%$ on average in reducing pursuit time. The code is available at \url{https://github.com/ANT-ITS/OARLM2I2}.
\end{abstract}


\section{Introduction}
With the development of Smart City, Intelligent Transportation System (ITS) \cite{ITS1} effectively leveraging the Internet of Vehicles (IoV) technology brings a profound impact on people’s lives \cite{IOV1, IOV3}. Multi-vehicle pursuit (MVP), a special and realistically meaningful problem in ITS, has been widely attracted. For example, the vehicle pursuit guideline \cite{newyork} has been published by the New York police department details the tactical operations to improve pursuit efficiency while cooperatively pursuing suspected vehicles.

\begin{figure*}[h]
\centerline{\includegraphics[width=\textwidth]{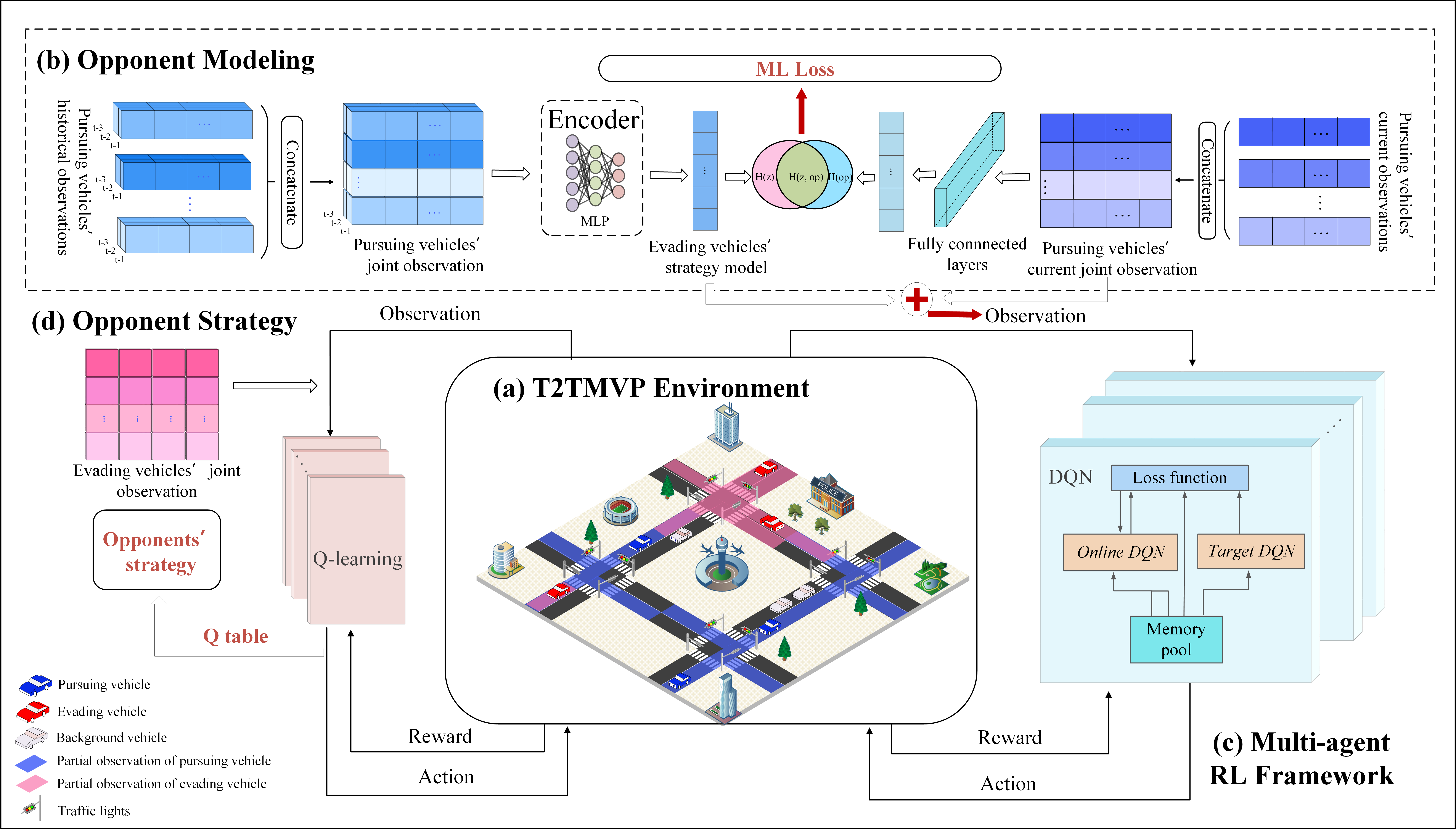}}
\caption{Overall Architecture of OARL$\text{M}^2\text{I}^2$. (a) Complicated urban traffic scene for T2TMVP problem, consisting of traffic lights, background vehicles, evading vehicles, and pursuing vehicles. (b) SEOJSM mechanism. This mechanism models the dynamic strategies of opponents assisted by mutual information-united loss. (c) Multi-agent reinforcement learning framework for pursuing agents. Each pursuing agent adopts DQN to make decisions with the assistance of the opponents joint strategy model. (d) State-sensitive joint dynamic strategy of opponents. Each evading agent leverages Q-learning to select actions with the highest Q-values.
}
\label{fig_1}
\end{figure*}

Essentially, the MVP problem can be modeled as pursuit-evasion game (PEG). In recent years, multi-agent reinforcement learning (MARL), showing significant advances in intelligent decision-making, has proven to be a fruitful method in PEG. Aiming at improving the cooperation between pursuers, \cite{ MARL3, MARL7} separately introduced curriculum learning and cross-task transfer learning in PEG. \cite{MARL6} proposed attention-enhanced reinforcement learning to address communication issues for multi-agent cooperation. As for homogeneous agents in MVP, \cite{ MARL5} proposed a transformer-based time and team reinforcement learning scheme. In addition to cooperation, some studies focus on the influence of opponents. \cite{MARL1} focused on predicting the future trajectory of the opponent to promote pursuit efficiency. However, these studies ignore the influence of the opponent’s strategy, especially when the opponent is characterized by a dynamic strategy which will bring extreme non-stationarity to the pursuit and thus increase the difficulty as well as randomness to a successful capture.

The opponent modeling method is integrated into MARL as a promising solution \cite{OPPO8} for building up the cognition of the opponent's dynamic strategy and alleviating the non-stationarity during the pursuit. In self-play scenarios, \cite{OPPO3} recursive reasons the opponent’s reactions to the protagonist’s potential behaviors and finds the best response. Targeting the non-stationarity brought by opponent’s changing behaviors, \cite{OPPO5} learned a general policy adaptive to changeable strategies. \cite{OPPO6} used policy distillation method to realize accurate policy detection and reuse in face of non-stationary opponents. \cite{OPPO7} learned low-level latent dynamics of the opponent, and leveraged the stability reward to stabilize the opponent strategy reducing the non-stationarity in tasks. However, the aforementioned methods suffer from a non-adaptation to the team-to-team multi-vehicle pursuit problem. On the one hand, state-of-the-art methods only focused on the two-player game and were difficultly adaptive to team-to-team competitions for that both generating and modeling complex strategies of opponents are challenging. On the other hand, the existed opponent modeling methods based on MARL is rarely applied to MVP scenario with complicated road structures and traffic restrictions.

This paper considers a team-to-team multi-vehicle pursuit problem (T2TMVP) in the complicated urban traffic scene. The evading vehicles adopt the pre-trained policy to choose the optimal actions rather than move randomly or in a fixed route, which is what we call dynamic strategies. The main target of this paper is allivating the non-stationarity brought by dynamic strategies of evading vehicles and further improve pursuit efficiency. For this purpose, an opponent-aware reinforcement learning via maximizing mutual information indicator (OARL$\text{M}^2\text{I}^2$) method is proposed to improve pursuit efficiency as shown in Fig. \ref{fig_1}. OARL$\text{M}^2\text{I}^2$ is equipped with the sequential encoding-based opponents joint strategy modeling (SEOJSM) mechanism to extract the joint features of dynamic strategies of evading vehicles based on Q-learning. Meanwhile, the DQN based pursuing vehicles implement efficient decision-making by leveraging the joint partial observation and the joint strategy model of evading vehicles, and the mutual information between them is served as an indicator to update the SEOJSM mechanism.
The main contributions of this paper are as follows:

1.	This paper models the team-to-team multi-vehicle pursuit (T2TMVP) problem in a complicated urban traffic scene. Two competitive teams, pursuing vehicle team and evading vehicle team, separately make flexible decisions according to intelligent dynamic strategies.

2.	This paper proposes an opponent-aware reinforcement learning via maximizing mutual information indicator (OARL$\text{M}^2\text{I}^2$) method to improve the pursuit efficiency for the T2TMVP problem. A sequential encoding-based opponents' joint strategy modeling (SEOJSM) mechanism is deliberately designed to assist in tackling the non-stationarity brought by dynamic strategies of evading vehicles.

3.	This paper leverages the novel mutual information-united loss to train our OARL$\text{M}^2\text{I}^2$. The mutual information-united loss comprehensively considers the effectiveness of decision-making network and opponents' joint strategy model.

The outline of this article is given as follows. Section \ref{s2} introduces the T2TMVP problem statement and problem instantiation. In Section \ref{s3}, the state-sensitive joint dynamic strategy of evading vehicles is introduced, and the SEOJSM mechanism is proposed. Section \ref{s4} details the deep Q-network for pursuing agents and the training process with the mutual information-united loss. Section \ref{s5} provides experiment settings and sufficient experiments to verify the effectiveness of the proposed OARL$\text{M}^2\text{I}^2$ method. Finally, conclusion and future work are presented in Section \ref{s6}.

\section{ T2TMVP Problem Statement and Instantiation} 
\label{s2}
In this section, we first state the T2TMVP problem. Then, we instantiate the T2TMVP problem as a partially observed Markov decision processes (POMDP).

\subsection{T2TMVP Problem Statement}
\begin{figure}[t]
\centering
\centerline{\includegraphics[width=\columnwidth]{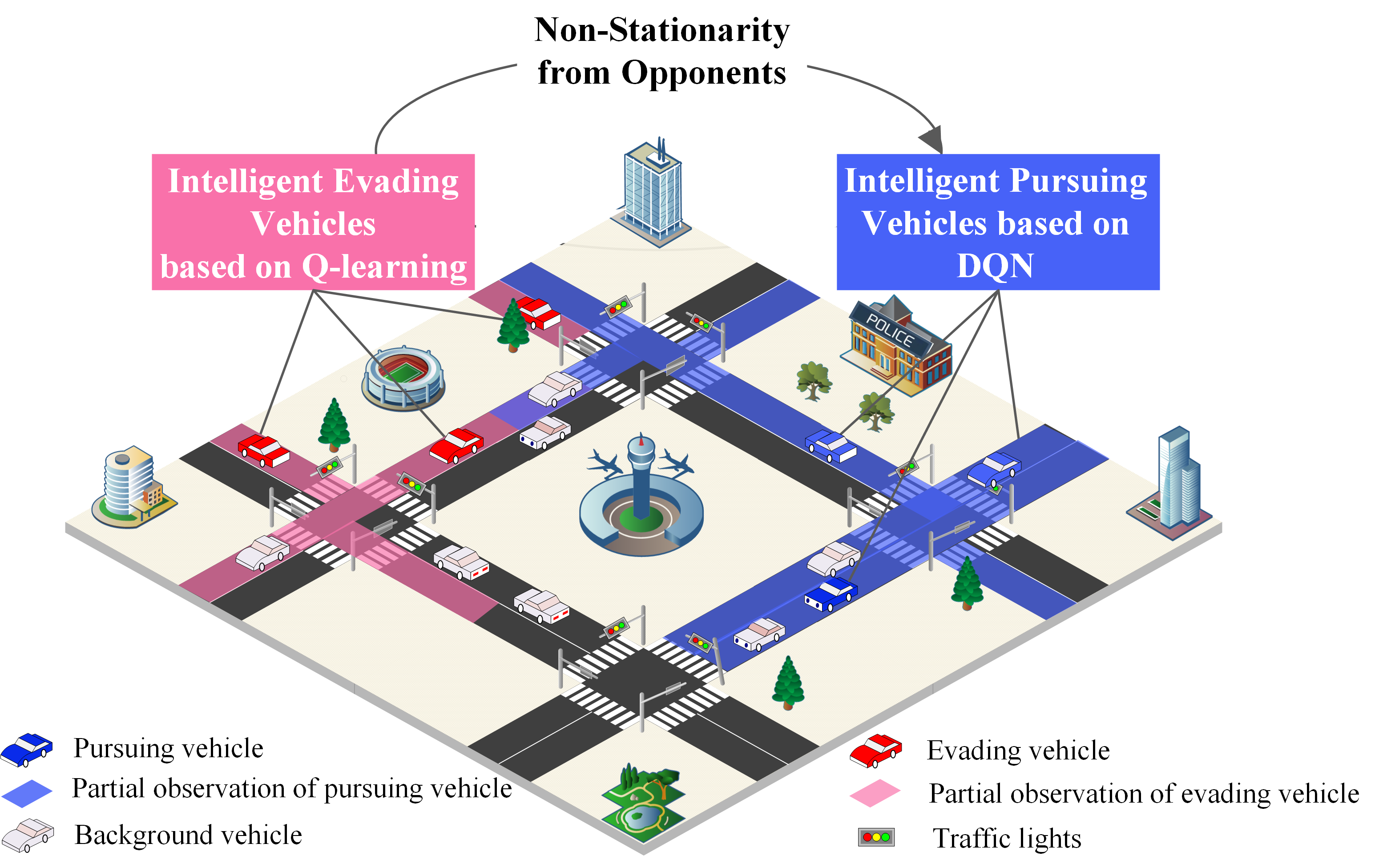}}
\caption{T2TMVP problem visualization. The Q-learning based evading vehicles and Deep Q-learning based pursuing vehicles compete in the complicated traffic scene.
}
\label{fig_2}
\end{figure}

This paper considers a team-to-team multi-vehicle pursuit (T2TMVP) problem in a complicated urban traffic scene as shown in Fig. \ref{fig_2}. Competition is the vital theme of T2TMVP, and two competitive teams of vehicles make intelligent decisions to separately accomplish their own goals. Different from the traditional MVP, in the T2TMVP problem, evading vehicles adopt the pre-trained policy to choose the optimal actions rather than move randomly or in a fixed route. As for the pursuing vehicles, the policy is constantly updated in the interactions with the environment. $I$ intersections and $L$ lanes form the structured bidirectional traffic topology. For lane $l \in \left\{1, 2, \ldots, L\right\}$, the adjacent lanes ahead are represented by $l_{rig}, l_{lef}, l_{str}$, whose subscripts means relative positions with lane $l$. In the complicated urban traffic scene, $B$ background vehicles exist similar to the real traffic scenario, and all vehicles are restricted to obey the following traffic rules in our simulation. 

(1)	Vehicles should follow traffic lights and drive in a single lane, turning is not allowed before reaching an intersection. 

(2)	Vehicles cannot exceed the speed limit, so both pursuing vehicles and evading vehicles are set the same acceleration $\mathop {ac}\nolimits_{\max}$ and the maximum speed permitted $\mathop v\nolimits_{\max}$. 

(3)	Collisions are considered when two vehicles get too close, the vehicle behind would decelerate at the maximum deceleration $\mathop {de}\nolimits_{\max}$ to prevent accidents.

In such a complicated urban traffic scene, an efficient pursuit is rather difficult. For one thing, the complex road structure and traffic regulations bring a lot of interference to the pursuit. For another, the non-stationarity caused by dynamic strategies of evading vehicles will impose extreme difficulties for pursuing vehicles learning optimal policies. Therefore, restricted by the complicated urban traffic scene, solving the non-stationarity issue caused by opponents is crucial for an efficient pursuit in the T2TMVP problem.

\subsection{T2TMVP Problem Instantiation}
In this paper, all vehicles, except for the background vehicles, make decisions based on the current partial observations restricted by the urban traffic scene. Therefore, the decision-making process of both pursuing vehicles and evading vehicles can be formulated as a partially observed Markov decision process (POMDP) define by a tuple $\left<S, O, A, P, R\right>$.  $s \in S$, $a \in A$ represents the global state space and action space. $o \in O$ is the partial observation of each agent. During the interaction with environment at time step $t$, each agent $k \in \left\{1, 2, \ldots, K\right\}$ chooses an action $a^{t}_{k}$ based on the obtained partial observation $o^{t}_{k}$ and forming the joint action $\mathbf{a}^{t}$. Then the environment generates the next state $s^{t+1}$ according to the state transition function $P\left(s^{t+1} \mid s, \mathbf{a}^{t}\right): S^{t} \times A^{t}_{1} \times A^{t}_{2} \times \ldots \times A^{t}_{K} \rightarrow S^{t+1}$. And reward $r^{t}_{k}$ as the feedback of action selection is given from the environment. The goal of each agent is to generate a optimal policy $\pi^{t}_{k}$  maximizing the discounted reward $R^{t}=\sum_{i=0}^{\infty} \gamma^{i} r^{t+i}$ and $\gamma \in [0,1]$ is the discount factor.

In the T2TMVP problem, the position of $N$ pursuing vehicles and $M$ evading vehicles are initialized randomly on the lanes. The goal of pursuing vehicles is to capture all evading vehicles in the shortest time possible, and the evading vehicles intend to escape accordingly. We consider a capture successful if, at any time step $t$ during the pursuit, the distance between an evading vehicle $m$ and at least one of the pursuing vehicles $n$ is less than a given collision radius $dis^{t}_{n,m}<dis_{cap}$. To be more realistic, the observations of pursuing vehicles $op^{t}_{n} \in OP, n \in \left\{1,2, \ldots, N\right\}$ and that of evading vehicles $oe^{t}_{m} \in OE, m \in \left\{1,2, \ldots, M\right\}$ are all restricted to be partial, and observations are shared within the homogeneous vehicles forming the joint observation $\mathbf{op}^{t} \in \mathbf{OP} $ and $\mathbf{oe}^{t} \in \mathbf{OE}$. When a vehicle encounters an intersection $Inter_{i}, i \in \left\{1,2, \ldots, I\right\}$, decision-making is needed. The evading vehicle $m$ adopts an action $a^{t}_{m} \in A$ according to pre-trained state-sensitive joint dynamic strategy $\mathbf{\pi}^{t}_{e}$  based on Q-learning. And the pursuing vehicle $n$ executes an action $a^{t}_{n} \in A$ through DQN against the non-stationary brought by opponents’ dynamic strategies. Empirically, in the process of driving, the observations encountered are generally limited, which is consistent with the limited state space of the Q-learning algorithm. We use the finite state-action pairs in Q-leaning to simulate the situation of adopting corresponding strategies for different observations in the driving process, which is what we call dynamic strategies. However, the state space of DQN is infinite and can not generate denumerable strategies for evading vehicles. For pursuing vehicles, we use the DQN algorithm to make decisions. On the one hand, DQN could make more refined decisions for the current observations. On the other hand, it is easy to compare with state-of-the-art algorithms.

\section{Opponent Modeling}
\label{s3}
This section first introduces the generating process of evading vehicles’ joint dynamic strategy based on Q-learning. Then, the SEOJSM mechanism is introduced.

\subsection{Joint Dynamic Strategy of Evading Vehicles}
\begin{figure}[h]
\centering
\centerline{\includegraphics[width=\columnwidth]{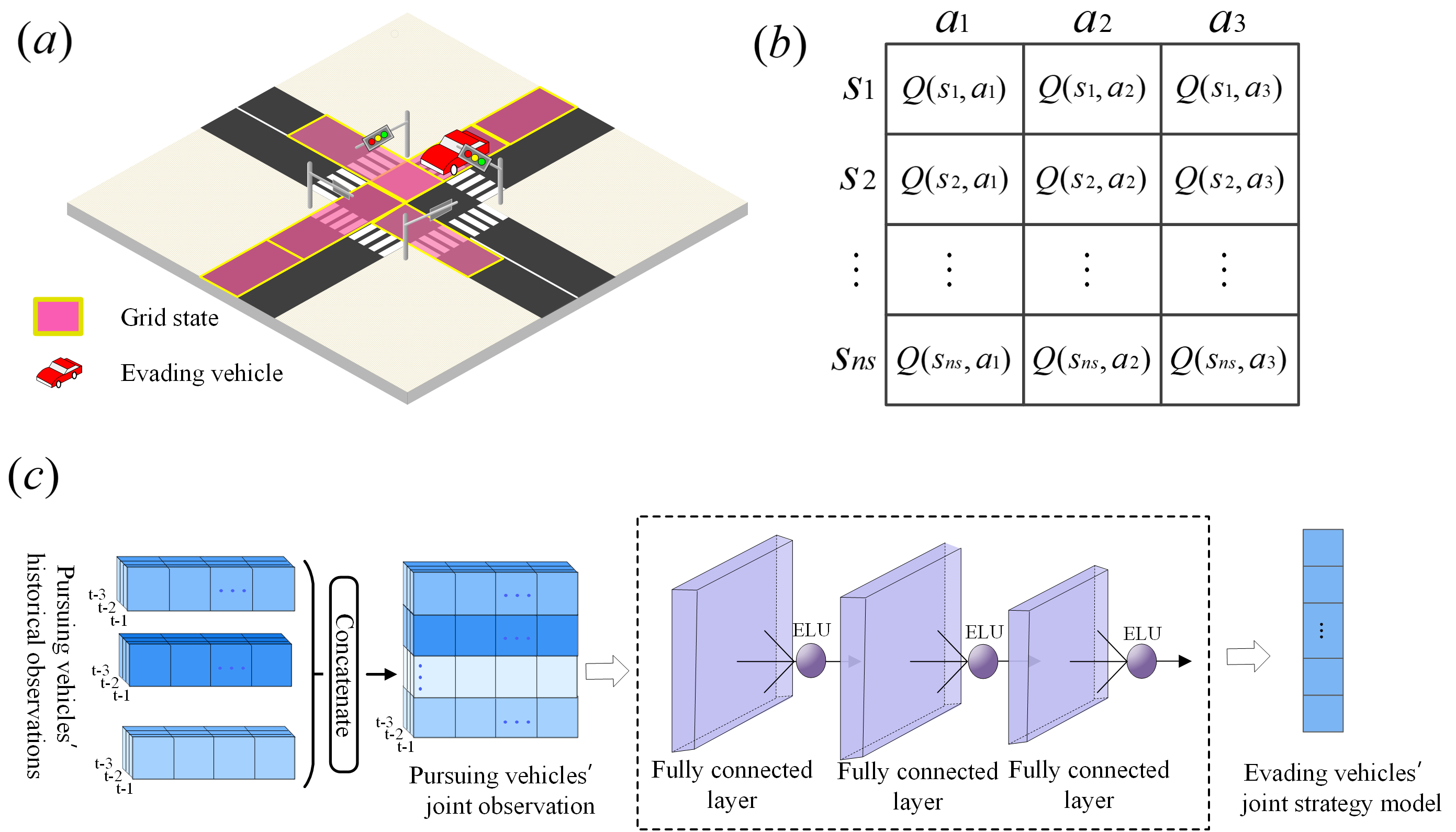}}
\caption{Opponent joint dynamic strategy generating and modeling. (a) Diagram of partial observation in the urban traffic scene. It describes the scope of partial observation and a cell state representation approach. (b) Opponent joint strategy generating process. Based on the Q-learning method, evading vehicles select the action according to the state-action pairs in the Q-table. (c) The SEOJSM mechanism, it leverages the historical joint observation of pursuing agents to generate the joint strategy model of opponents.
}
\label{fig_3}
\end{figure}

In the traditional opponent modeling methods, an evading vehicle tends to choose a strategy from a few preset fixed strategies based on the current observation. The previous adversary modeling is in the scenario of two-agent and the state space of decision-making is less, but the state space of decision making increases exponentially in the T2TMVP problem, therefore the preset strategies are not enough to make effective decisions. However, due to the huge state space of evading vehicles in the continuous scene of complicated urban traffic, the above evading strategies are not applicable. Moreover, preset strategies focus on dealing with a few simple cases and it is difficult for them to make cooperative decisions from the perspective of a single agent. To this end, this paper delicately designs a novel strategy-generating approach for multiple evading vehicles in the T2TMVP problem as shown in Fig. \ref{fig_3} (a), (b).

Q-learning is one of the effective algorithms of reinforcement learning. It introduces the mapping Q-table between the state-action pairs and the corresponding estimated future rewards into the action selection process of an agent. According to the current state $s$, an agent selects the action $a$ following policy $\pi$. If the state-action pair is not contained in the Q-table, then the action-utility function $Q(s, a)$ will be updated.  Inspired by this, this paper leverages the Q-learning method to generate the joint dynamic strategy of evading vehicles. Thus, evading vehicles enable intelligent executions based on the current states.

This paper uses a cell state representation approach to model the state space of evading vehicles. We divide each lane in an agent’s visual field into two cells as shown in Fig. \ref{fig_3} (a). For an evading agent $m$ on lane $l$, the lane $l$ and the connected lane ahead $l_{str}$ are fully observed, and the connected lanes $l_{rig}$, $l_{lef}$ in the lateral visual field are restricted to half of the lane length for realism. Therefore, lanes in the visual field are divided into six cells $\left\{Cel^{1}_{l}, Cel^{2}_{l}, Cel^{1}_{l_{str}}, Cel^{2}_{l_{str}}, Cel^{1}_{rig}, Cel^{1}_{lef}\right\}$, as shown in Fig. \ref{fig_3} (a). Moreover, we let the partial observation $oe^{t}_{m}$, at time step $t$, consist of the number of pursuing agents on every cell in the visual field, forming $oe^{t}_{m}= [num^{t}_{Cel^{1}_{l}}, num^{t}_{Cel^{2}_{l}}, num^{t}_{Cel^{1}_{l_{str}}}, num^{t}_{Cel^{2}_{l_{str}}}, num^{t}_{Cel^{1}_{rig}},\\
num^{t}_{Cel^{1}_{lef}}]$.

In the training process of evading agents, with the same complicated urban traffic scene as mentioned before, pursuing agents are set to randomly move, and evading agents choose the optimal actions using the Q-learning, as shown in Fig. \ref{fig_3} (b). Decision-making will only take place when vehicles reach an intersection. Therefore, three actions, going straight, turning right, and turning left, consist of the action space $\mathbf{A}$. At time step $t$, via cooperation among the evading vehicles, the partial observations $\left\{oe^{t}_{1}, oe^{t}_{2}, \ldots, oe^{t}_{M}\right\}$ form the joint observation $\mathbf{oe}^{t}$. The evading agent $m$ selects an action $a^{t}_{m} \in \mathbf{A}$  and performs it condition on the current joint observation $\mathbf{oe}^{t}$. After receiving the environment reward $r^{t}_{m}$, $Q$ value is updated based on the following Bellman equation:

\begin{equation}
\begin{aligned}
Q(\mathbf{oe}^{t}, a^{t}_{m}) 
\leftarrow Q(\mathbf{oe}^{t}, a^{t}_{m})
+\alpha[r^{t}_{m}&+\gamma \max _{a^{t+1}} Q\left(\mathbf{oe}^{t+1}, a^{t+1}\right)\\
&-Q(\mathbf{oe}^{t}, a^{t}_{m})],
\end{aligned}
\end{equation}
where $\alpha$ is the learning rate, and $\gamma$ is the discount factor. During the training process, Q-table is updated constantly with new state-action pairs the evading vehicles will encounter, and the Q-value corresponding to a state-action pair is replaced by a higher one. Each evading vehicle selects the action with the highest Q-value based on the current joint partial observation. That is exactly the state-sensitive joint dynamic strategy of evading vehicles which guides the competition with pursuing vehicles. It provides a dynamic strategy based on the current observation for evading vehicles but also considers the team tactics.

\subsection{Sequential Encoding-Based Opponents' Joint Strategy Modeling Mechanism}
This paper proposes the SEOJSM mechanism to learn the joint strategy model of evading vehicles, as shown in Fig. \ref{fig_3} (c). We leverage multi-layer perceptron (MLP), consisting of multiple fully connected layers and the activation function ELU, serves as an encoder to build up the cognition of evading vehicles' joint dynamic strategy. 

In the T2TMVP problem, due to cooperation existing in both pursuing team and evading team, clear position representation is important to obtaining effective information and further improving pursuit efficiency. At time step $t$, the position representation of vehicle $k$ is given by $Loc_{k}=\left[Emb_{l}, Emb_{l_{str}}, Emb_{l_{lef}}, Emb_{l_{rig}}, dis^{t}_{k,l} \right]$. Here, $Emb_{l}$ is the one-hot encoding of the lane $l$ on which the agent $k$ is located, $Emb_{l_{str}}$, $Emb_{l_{lef}}$, $Emb_{l_{rig}}$ represent the one-hot encoding of the lanes the agent $k$ is access to by executing going straight, turning left and turning right, respectively. $dis^{t}_{k,l}$ is the distance between the position of agent $k$ and the start of the located lane $l$. Hence, the partial observation of the pursuing vehicle $n$ is presented as $op^{t}_{n}=\left[\mathbf{Loc}_{Mv^{t}_{n}}, \mathbf{Loc}_{N}, adj\right]$. Here, $\mathbf{Loc}_{Mv^{t}_{n}}$ represents the joint position representations of $Mv^{t}_{n}$ evading vehicles in the visual field, $\mathbf{Loc}_{N}$ means the joint position representation of all pursuing vehicles, and $adj$ represents the road topology information as the extra information. Therefore, the joint partial observation of pursuing vehicles is denoted as $\mathbf{op}^{t}=\left[\mathbf{Loc}_{N}, \mathbf{Loc}_{Mv^{t}_{total}}, adj\right]$, consists the joint position representations of all pursuing vehicles $\mathbf{Loc}_{N}$, the joint position representation $\mathbf{Loc}_{Mv^{t}_{total}}$ of the total $Mv^{t}_{total}$ evading vehicles, and the road topology information $adj$, where $Mv^{t}_{total}=\sum_{n=1}^{N} Mv^{t}_{n}$.

In the SEOJSM mechanism, we feed the it with the joint historical partial observations in the past $h$ time steps of all pursuing agents $\left[\mathbf{op}^{t-3}, \mathbf{op}^{t-2}, \mathbf{op}^{t-1}\right]$. And the joint strategy model of all evading vehicles $\mathbf{\pi}^{t}_{e}$ is output, realizing the strategy cognition building up towards evading vehicles.

Our key insight is that building evading vehicles’ joint strategy model only using partial observation is a concise, realistic, and effective method. Knowing the likely strategy of opponents influences a pursuing vehicle’s beliefs over environmental states and thus informs its planning of future actions. The reason for generating a joint strategy model instead of separate strategy models for every evading agent is that evading team also works collaboratively, thus pursuing agents can not only infer the single evading agent’ strategy but also recognize the tactical of the whole team from the joint strategy model.

\section{Deep Q-network with united loss for Pursuing Vehicles} 
\label{s4}
In this section, we first introduce the deliberately designed ingredients of deep Q-networks for pursuing vehicles in the T2TMVP problem. Then we illustrate the training process with the mutual information-united loss.

\subsection{Opponent-Aware Deep Q-networks for Pursuing Vehicles}
DQN, as an upgraded version of Q-learning, is widely used in intelligent decision-making with discrete action space. This paper leverages DQN to provide decision-making for each pursuing vehicle in the T2TMVP problem. Based on Q-learning, DQN sets a neural network to estimate the current action-utility function $Q$ and outputs the Q-value of each action condition on the current state. The DQN-based agent implements optimal decision-making by selecting the action with the highest Q-value. In this paper, we adapt DQN to the T2TMVP problem with the following paradigm setting of reinforcement learning, including the state representation, the action space, and the reward structure.

In the T2TMVP problem, the evading vehicles conduct flexible decisions according to the current state making them elusive for pursuing vehicles, thus the dynamic strategy of evading vehicles brings extreme non-stationarity to the pursuit task. In this paper, we feed two parts of input into DQN for efficient decision-making, consisting of the joint partial observation of pursuing vehicles and the joint strategy model of evading vehicles. As described in Section \ref{s3}, the joint partial observation of pursuing vehicle at time step $t$ is represented as $\mathbf{op}^{t}=\left[\mathbf{Loc}_{N}, \mathbf{Loc}_{Mv^{t}_{total}}, adj\right]$. And the joint strategy model of evading vehicles output by the SEOJSM mechanism is $\mathbf{\pi}^{t}_{e}$. We concatenate the above two parts forming the state $s^{t}=\left[\mathbf{op}^{t}, \mathbf{\pi}^{t}_{e}\right]$, then leverage the concatenation jointly predicting the Q-value.

In the decision-making process of DQN, the neural network eventually outputs the Q-value for each action indicating the maximized future rewards if implementing the action. In the T2TMVP problem, action execution takes place only when vehicles reach the intersection. Therefore, the action space is set as the general intuition $\mathbf{A} = \left\{a_{1}, a_{2}, a_{3}\right\}$, containing going straight $a1$, turning left $a2$, and turning right $a3$.

At each time step $t$, the pursuing vehicle individually receives a reward designed to incentive the capture of evading vehicles. For pursuing vehicle $n$, the formulation of the reward function $r_{n}$ is as follows:

\begin{equation}
\begin{aligned}
r^{t}_{n} = - \lambda \sum_{m=1}^{Mv^{t}_{n}} \left(dis^{t}_{n,m} – dis^{t-1}_{n,m}\right) -c +  r^{t}_{n, task}.
\end{aligned}
\end{equation}
Here, the reward function is deliberately designed in three aspects. The distance-based reward $r^{t}_{n, dis}=- \lambda \left(dis^{t}_{n,m} – dis^{t-1}_{n,m}\right)$ is responsible impelling pursuing vehicle $n$ to reduce the distance with the nearest evading vehicle $m$ in the visual field and continuously move towards the opponent. $\lambda$ is the distance-based reward factor. To incentive faster pursuit, a time-based reward $r^{t}_{n, time}=-c$ works by imposing a negative reward $c$ every time step until completing a successful pursuit. When an evading agent is captured, all pursuing agents will be given a task-based reward $r^{t}_{n, task}$ indicating the effectiveness of cooperation.

\subsection{Training with Mutual Information-United Loss}
The training regime for OARL$\text{M}^2\text{I}^2$ is identical to the original DQN. 
DQN adopts the double-network structure. The online network with the parameter $\theta_{Q}$ approximates the $Q\left(s^{t}_{n},a^{t}_{n}\right)$ and update the parameter $\theta_{Q}$, and the target network with the parameter $\theta^{\prime}_{Q}$ calculates the Q-target $y\left(t\right)=r^{t}_{n}+\max_{{a}^{t+1}_{n}}Q\left(s^{t+1}_{n},a^{t+1}_{n} \mid \theta^{\prime}_{Q}\right)$ and updates the parameter $\theta^{\prime}_{Q}$ with $\theta_{Q}$ at regular intervals. The double-network structure avoids the instability caused by updating the Q-function while obtaining the Q-value, thus making the update smooth and accelerating the convergence of the algorithm.

In the original DQN, the Q-function can be learned by minimizing the following MSE loss function between the Q-target $y\left(t\right)$ and $Q\left(s^{t}_{n}, a^{t}_{n}\right)$. The expectation term is approximated by sampling a batch uniformly at random from a replay buffer containing past transition tuples. The original optimizing objective is as follows:
\begin{equation}
\begin{aligned}
L_{1}(\theta_{Q})=\mathbb{E}\left[\left(Q\left(s^{t}_{n}, a^{t}_{n} \mid \theta_{Q}\right)- y\left(t\right)\right)^{2}\right].
\end{aligned}
\end{equation}

Noting that the SEOJSM mechanism cannot guarantee the anticipation of opponents’ strategies models, we introduce an explicit regularization to guide the modeling process. Mutual information (MI) measures the information shared by two variables, i.e. the degree to which the uncertainty of variable $X$ is reduced by obtaining variable $Y$. The similarity of $X$ and $Y$ would be improved if the mutual information increases.

\begin{algorithm}[h]\label{alg_1}
            \caption{Training process of OARL$\text{M}^2\text{I}^2$}
            \LinesNumbered
            Initialize replay buffer $R$\;
            Initialize action-value function $Q$ with random weights $\theta_{Q}$\;
            Initialize target action-value function $\hat{Q}$ with random weights $\theta^{\prime}_{Q} \leftarrow \theta_{Q}$\;
            \For{$episode=1$ to $Ep$}{
            Receive initial state $\mathbf{s}^{1}=\left[\mathbf{op}^{t},\mathbf{\pi}^{1}_{e}\right]$\;
            Initialize and store $\mathbf{\pi}^{1}_{e}$ in observation pool $H$\;
            \For{$time step=1$ to $T$}{
            \For{each pursuing agent $n$}{
            Choose action $a^{t}_{n}=\pi\left(\mathbf{op^{t}} \mid \theta_{Q} \right)$ and exploration\;
            }
            Take action $\mathbf{a}^{t}=\left[a^{t}_{1}, a^{t}_{2}, \ldots, a^{t}_{N}\right]$\; 
            Get reward $\mathbf{r}^{t}=\left[r^{t}_{1}, r^{t}_{2}, \ldots, r^{t}_{N}\right]$\;
            Obtain new joint partial observation $\mathbf{op}^{t+1}$\;
            Store joint partial observation $\mathbf{op}^{t+1}$ in observation pool $H$\;
            Feed the SEOJSM mechanism with the latest $h$-step historical data in $H$ and output opponents' joint strategy model $\mathbf{\pi}^{t+1}_{e}$\;
            Form state $\mathbf{s}^{t+1}_{n}$ with $\mathbf{op}^{t+1}$ and $\mathbf{\pi}^{t+1}_{e}$ and form $\mathbf{s}^{t+1}=\left[s^{t+1}_{1},s^{t+1}_{2}, \ldots, s^{t+1}_{N}\right]$\;
            Store transition $\left(s^{t}_{n},a^{t}_{n},r^{t}_{n} ,s^{t+1}_{n}\right)$ in $R$\;
            Sample a random mini-batch of transitions $ (\mathbf{s}, \mathbf{a}, \mathbf{r}, \mathbf{s}^{\prime}) $ from $R$\;
            Update $\theta_{Q}$ by minimizing the loss Equation(\ref{alg_5})\;
            Update the parameters of target action-value function $\theta^{\prime}_{Q} = \theta_{Q}$ with period $C$\;
            $\mathbf{s}^{t}=\mathbf{s}^{t+1}$, $\mathbf{\pi}^{t}_{e}=\mathbf{\pi}^{t+1}_{e}$
            }
            }
\end{algorithm}

Concerning the T2TMVP problem, the pursuing vehicles are eager for the evading vehicles’ escaping strategies at the next intersection to improve pursuit efficiency. As such, the quality of the opponents’ strategy model depends on whether it can accurately infer the opponents' next strategy. Taking this cue, this paper optimizes the opponent modeling model by maximizing the mutual information between the evading vehicles' joint strategy model $\pi^{t}_{e}$ and pursuing vehicles' joint observations $\mathbf{op}^{t}$. The calculating formulation is as follows:
\begin{equation}
\begin{aligned}
I(\mathbf{op}^{t}; \pi^{t}_{e})&=H\left(\pi^{t}_{e}\right)-H\left(\pi^{t}_{e} \mid \mathbf{op}_{t}\right)\\
&=\sum_{\mathbf{op}^{t}, \pi^{t}_{e}} p(\mathbf{op}^{t}, \pi^{t}_{e}) \log \frac{p(\mathbf{op}^{t}, \pi^{t}_{e})}{p(\mathbf{op}^{t}) p(\pi^{t}_{e})}.
\end{aligned}
\end{equation}
If the knowledge of entropy $H(\pi^{t}_{e})=$ does not provide any information about the entropy $H((\mathbf{op}_{t})$, the mutual information would become zero, which means the failure of the opponent modeling model. Therefore, the overall optimizing objective is:
\begin{equation}
\label{alg_5}
\begin{aligned}
L(\theta) = L_{1}(\theta_{Q}) - I(\mathbf{op}^{t}; \pi^{t}_{e}).
\end{aligned}
\end{equation}

The overall training process of OARL$\text{M}^2\text{I}^2$ is demonstrated in Algorithm \ref{alg_1}. At the beginning of each episode, the joint partial observation of all pursuing vehicles $\mathbf{op}^{1}$ and the joint strategy model of evading vehicles $\mathbf{\pi}^{1}_{e}$ are initialized. At each time step $t$, each pursuing vehicle $n$ select an action $a^{t}_{n}$ according to the current policy with $1-\epsilon$ probability, and execute random choice with $\epsilon$ probability. The immediate reward $r^{t}_{n}$ will be provided by the environment. And new partial observation $op^{t+1}_{n}$ is received forming the joint partial observation $\mathbf{op}^{t+1}$, which is stored in the observation pool $H$. Then, the SEOJSM mechanism takes $h$-step joint partial observation $\left[\mathbf{op}^{t-2}, \mathbf{op}^{t-1}, \mathbf{op}^{t} \right]$ as input to output the joint strategy model of evading vehicles $\mathbf{\pi}^{t+1}_{e}$ which forms the state $s^{t+1}_{n}$ with the joint partial observation $\mathbf{op}^{t}$. The transition $\left(s^{t}_{n},a^{t}_{n},r^{t}_{n} ,s^{t+1}_{n}\right)$ is then stored in the replay buffer $R$. Finally, samples are selected from the replay buffer $R$ to update all networks.

\section{Performance Comparison and Analysis}
\label{s5}
This section discusses the performance of our OARL$\text{M}^2\text{I}^2$ method in the complicated urban traffic scene. First, we introduce the experience setting of the simulation. Then, we compare the proposed OARL$\text{M}^2\text{I}^2$ with state-of-the-art RL methods DQN, PPO, and QMIX, as well as OARL$\text{M}^2\text{I}^2$ inaccessible with the road topology information, in the long-term reward, discounted reward, convergence, and optimal performance four aspects. The result analysis is detailed in subsections.

\subsection{Simulation Settings}
To train and evaluate our OARL$\text{M}^2\text{I}^2$, we propose T2TMVP in the complicated urban traffic setting on SUMO. We construct an urban traffic scene with $3\times3$ interactions and $48$ lanes, where background vehicle flows follow the pre-set routes. $2$ evading vehicles and $4$ pursuing vehicles are competing in the scene. The parameters concerning the pursuit scenario are demonstrated in the upper part of Table \ref{tab1}. 

\begin{table}[h]
\centering
\caption{Parameter settings}
\label{tab1}
\renewcommand\arraystretch{1}
\begin{tabular}{c|c}
\hline
Parameters                                         & Value  \\ \hline
number of pursuing vehicles $N$                    & 6      \\
number of evading vehicles $M$                     & 3      \\
number of background vehicles $B$                  & 50     \\
capture radius $dic_{cap}$                         & 5 m     \\
maximum speed $\mathop v\nolimits_{\max}$           & 20 m/s    \\
maximum acceleration $\mathop {ac}\nolimits_{\max}$ & 0.5 m/$\text{s}^2$   \\
maximum deceleration $\mathop {de}\nolimits_{\max}$ & -4.5 m/$\text{s}^2$ \\ \hline
historical joint observation length $h$            & 3 time step    \\
reward per time step $c$                           & -0.2   \\
reward for capture $r_{task}$                      & 10     \\
weight of distance-based reward $\lambda$          & 2      \\
greedy factor $\epsilon$                           & 0.05   \\ 
discount factor $\gamma$                           & 0.95   \\ \hline

\end{tabular}
\end{table}

The SEOJSM mechanism is designed with three fully-connected hidden layers with 128 units. ELU activation function is used for each hidden layer. In Deep Q-network, the maximum steps of an episode are restricted. The batch size is set as 32, the learning rate is set as 0.001. The replay buffer capacity is 10000 and the Adam optimizer is used during training. The GPU used in training is NVIDIA Tesla T4. The parameters of OARL$\text{M}^2\text{I}^2$ is shown in the below part of Table \ref{tab1}.

\subsection{Ablation Analysis}
This subsection analyzes the effectiveness of the SEOJSM mechanism and the mutual information-united loss from several perspectives. 

\begin{figure}[h]
    \centering
    \centerline{\includegraphics[width=0.9\columnwidth]{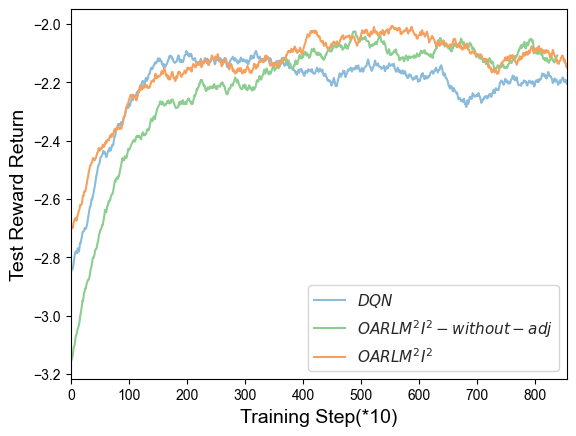}}
    \caption{Undiscounted reward comparison in test.}
    \label{result2}
\end{figure}
\begin{table}[h]
\centering
\caption{Optimal Performance}
\label{table2}
\renewcommand\arraystretch{1} 
\resizebox{\columnwidth}{!}{ 
\begin{tabular}{r|cccc}
\hline
\multicolumn{1}{c|}{} & \begin{tabular}[c]{@{}c@{}}best undisco-\\ unted return\end{tabular} & \begin{tabular}[c]{@{}c@{}}best promotion\\comparsison\end{tabular} & \begin{tabular}[c]{@{}c@{}}best \\ time step\end{tabular} & \begin{tabular}[c]{@{}c@{}}best promotion\\comparsison\end{tabular} \\ \hline
OARL$\text{M}^2\text{I}^2$             & \textbf{-1.347} & —                & \textbf{236.3} & —                \\
OARL$\text{M}^2\text{I}^2$-without-adj & -1.527          & 11.79\%          & 298.4          & 20.81\%          \\
DQN               & -1.411          & 4.53\%           & 327.4          & \textbf{27.82\%} \\
PPO               & -1.639          & \textbf{17.81\%} & 289.5          & 18.38\%          \\
QMIX              & -1.375          & 2.03\%           & 287.4          & 17.78\%          \\ \hline
\end{tabular}
} 
\end{table}

We first analyze the optimal performance including the best undiscounted return $R_{t}=\sum_{i=0}^{\infty}r_{t+i}$ and the minimum time steps finishing the pursuit, as shown in Table \ref{table2}. The best undiscounted reward represents the overall performance related to three aspects: time, distance and task. The best time step is the embodiment of pursuit efficiency. The proposed OARL$\text{M}^2\text{I}^2$ separately realizes 4.54$\%$ and 27.83$\%$ outperforming DQN on the best undiscounted reward and best pursuit time steps, respectively. The convincing results indicate that the SEOJSM mechanism and the mutual information-united loss jointly boost the pursuit performance in both reward and efficiency perspectives. In particular, the OARL$\text{M}^2\text{I}^2$ without road topology information surprisingly achieves 8.86$\%$ faster pursuit than DQN, despite being inferior to OARL$\text{M}^2\text{I}^2$, which confirms that the OARL$\text{M}^2\text{I}^2$ enable better ability dealing with complex pursuit task equipped with the joint strategy model output by SEOJSM mechanism.

The undiscounted reward enables an intuitive description of the direct feedback from the environment in training. In this part, we analyze the ablation experiment based on the undiscounted return as shown in Fig. \ref{result2}. Compared with DQN, OARL$\text{M}^2\text{I}^2$ combined with the SEOJSM mechanism and mutual information-united loss, which are considered inseparable in this paper. Depicted as Fig. \ref{result2}, OARL$\text{M}^2\text{I}^2$ outperforms DQN in general. Especially in the training periods of beginning and convergent, the superior performance indicates the SEOJSM mechanism and mutual information-united loss play an important role in the initial exploration and final performance. Moreover, the OARL$\text{M}^2\text{I}^2$-without-adj eventually achieves the competitive performance with OARL$\text{M}^2\text{I}^2$ proving the effective assistance of OARL$\text{M}^2\text{I}^2$ in pursuing vehicles' decision-making process despite in the information-inaccessible situation.

\begin{figure}[t]
    \centering
    \centerline{\includegraphics[width=0.9\columnwidth]{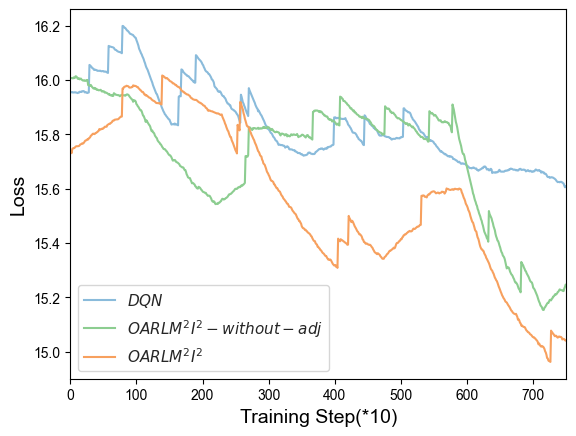}}
    \caption{Convergence comparison during training process.}
    \label{result3}
\end{figure}

Convergence is a vital factor in measuring the efficiency of an algorithm. As depicted in Fig. \ref{result3}, all experimental methods present a good convergence trend in our scenario. It is obvious that the proposed OARL$\text{M}^2\text{I}^2$ method achieves remarkable superiority over DQN. This confirms that the mutual information-united loss could assist in converging to a great extent. It is reasonable to infer that OARL$\text{M}^2\text{I}^2$ could generate effective policies for pursuing vehicles more quickly by inference intention of opponents’ strategy to decrease the non-stationarity. We also compare the loss of OARL$\text{M}^2\text{I}^2$ inaccessible with road topology information with OARL$\text{M}^2\text{I}^2$ and DQN. In the conspicuous training period from around step 300*10 to 600*10, the loss of OARL$\text{M}^2\text{I}^2$-without-adj soars and loses the original advantage. This could be interpreted as the exploration based on $\epsilon$ greedy strategy, the action space sampling increases resulting in large fluctuations in the descent. But as shown in the final result, the convergence of OARL$\text{M}^2\text{I}^2$-without-adj shows a little difference with OARL$\text{M}^2\text{I}^2$. It indicates that OARL$\text{M}^2\text{I}^2$ without topology information need time to constantly establish the cognition of the environment, and eventually, OARL$\text{M}^2\text{I}^2$ can suppress the uncertainty of complicated environment effectively via competitive decision-making.

\subsection{Comparison among Algorithms}
This paper compares the optimal performance of each algorithm in the best undiscounted return and the minimum time steps finishing the pursuit as shown in Table \ref{table2}. Analyze from the specific data, the best undiscounted return is 2.03$\%$ higher than QMIX with the second performance and 17.81$\%$ higher than PPO with the worst performance.  In the absence of road topology information, OARL$\text{M}^2\text{I}^2$ could still realize the 7.33$\%$ advantage over PPO. Thus, inference on the opponents’ dynamic strategy is vital in addressing the non-stationary pursuit task. In terms of the pursuit efficiency, our proposed OARL$\text{M}^2\text{I}^2$ has made significant advances with 18.83$\%$ and 17.78$\%$ outperforming PPO, and QMIX separately. And the proposed  OARL$\text{M}^2\text{I}^2$ outperforms other algorithms by 21.48$\%$ on average. In general, OARL$\text{M}^2\text{I}^2$ outperforms other algorithms by achieving the highest discounted return and accomplishing the pursuit in the shortest time. From the above analysis, the superiority of the proposed OARL$\text{M}^2\text{I}^2$ indicates the effectiveness of alleviating the non-stationarity brought by the dynamic strategy of opponents through the SEOJSM mechanism.

\begin{figure}[t]
    \centering
    \centerline{\includegraphics[width=0.9\columnwidth]{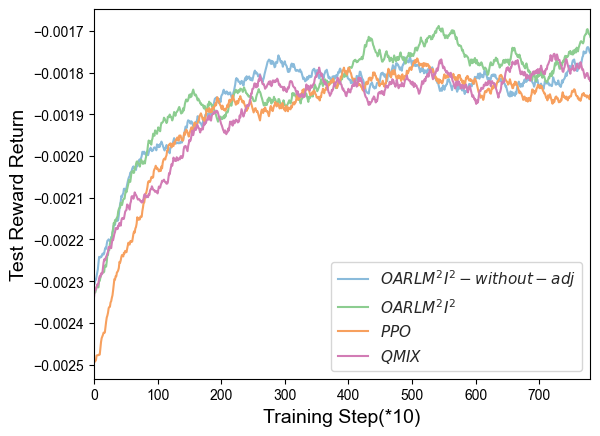}}
    \caption{Discounted reward comparison in test for algorithms. OARL$\text{M}^2\text{I}^2$-without-adj refers to removing the road topology information from the state.}
    \label{result1}
\end{figure}

More intuitively, this paper compares the discounted reward of the proposed OARL$\text{M}^2\text{I}^2$ and the state-of-the-art methods of PPO and QMIX to analyze the long-term performance as shown in Fig.\ref{result1}. The discount reward, $R_{t}=\sum_{i=0}^{\infty} \gamma^{i} r_{t+i}$, $\gamma \in [0,1] $, which can avoid the algorithm falling into local optimization and form a long-term policy, is a great index comparing the long-term performance. The OARL$\text{M}^2\text{I}^2$ converges to the highest reward, followed by the QMIX and PPO. It is worth noting that, although the reward setting is rather harsh for the high proportion of the time-based reward, it is obvious that OARL$\text{M}^2\text{I}^2$ presents an impressive advance at first. At the beginning of the pursuit, pursuing vehicles obtain a little experience towards the opponents where uncertainty reaches the highest, thus the competitive performance indicates that OARL$\text{M}^2\text{I}^2$ can cope with the non-stationarity of opponents to enhance the decision-making and effective policy-generating. 

\section{Conclusion and Future Works}
\label{s6}
This paper focuses on the T2TMVP problem in a complicated traffic scene with background vehicle flows and traffic lights. We propose an opponent-aware reinforcement learning via maximizing mutual information indicator (OARL$\text{M}^2\text{I}^2$) method to improve pursuit efficiency by tackling the non-stationary brought by opponents’ dynamic strategy. A SEOJSM mechanism is proposed to assist the decision-making process of pursuing vehicles by building up the cognition of evading vehicles’ dynamic strategy. Moreover, this paper proposes the mutual information-united loss synchronously update the SEOJSM mechanism and DQN-based multi-agent decision-making model to accelerate the convergence of OARL$\text{M}^2\text{I}^2$. Finally, we verify the OARL$\text{M}^2\text{I}^2$ method in a simulated complicated traffic scene based on SUMO. Extensive experiments demonstrate our approach outperforms other baselines by 21.48$\%$ on average in reducing pursuit time and presents better convergence. Our future works mainly focus on exploring more complex scenarios, such as larger traffic scenes, and different ratios of pursuing vehicles and evading vehicles. Another interesting direction lies in adversarial reinforcement learning, i.e. training the pursuing vehicles and evading vehicles simultaneously, which will impose challenges by introducing more non-stationarity.

\bibliographystyle{unsrt}
\bibliography{9772}

\end{document}